\documentclass[fleqn,10pt]{wlscirep}
\title{Signatures of the Kondo effect in VSe$_2$}

\author[1,*]{Sourabh Barua}
\author[1]{M. Ciomaga Hatnean}
\author[1]{M. R. Lees}
\author[1,*]{G. Balakrishnan}
\affil[1]{Department of Physics, University of Warwick, Coventry, CV4 7AL, United Kingdom}

\affil[*]{S.Barua.1@warwick.ac.uk, G.Balakrishnan@warwick.ac.uk}

\begin{abstract}
VSe$_2$ is a transition metal dichaclogenide which has a charge-density wave transition that has been well studied. We report on a low-temperature upturn in the resistivity and, at temperatures below this resistivity minimum, an unusual magnetoresistance  which is negative at low fields and positive at higher fields, in single crystals of VSe$_2$. The negative magnetoresistance has a parabolic dependence on the magnetic field and shows little angular dependence. The magnetoresistance at temperatures above the resistivity minimum is always positive. We interpret these results as signatures of the Kondo effect in VSe$_2$.  An upturn in the susceptibility indicates the presence of interlayer V ions which can provide the localized magnetic moments required for scattering the conduction electrons in the Kondo effect. The low-temperature behaviour of the heat capacity, including a high value of $\gamma$, along with a deviation from a Curie-Weiss law observed in the low-temperature magnetic susceptibility, are consistent with the presence of magnetic interactions between the paramagnetic interlayer V ions and a Kondo screening of these V moments.\end{abstract}
\begin{document}

\flushbottom
\maketitle
%
%
\thispagestyle{empty}

\section*{Introduction}

Transition metal dichalcogenides (TMDCs) are compounds of the form $TX_2$ made up of the transition metals ($T$) and elements of the chalcogen group ($X = $~S, Se, Te). Most of these compounds are layered but some non-layered materials are also formed by transition metal elements from Group VII and Group VIII\cite{Wilson1969}. The crystal structure of the layered materials is made by repetition of units consisting of a layer of the metal atoms sandwiched between two layers of chalcogen atoms and in each of these layers the atoms are arranged hexagonally. The bonding between the layers in a unit is covalent but the chalcogen atoms of two different layers are held by weak van der Waal forces and hence these materials cleave very easily perpendicular to the $c$-axis. Different polytypes of the same compound can exist depending upon how the chalcogen and the metal layer are stacked. A stacking of the AbA type leads to a trigonal prismatic coordination for the metal atom while a stacking of the AbC type leads to an octahedral coordination. The octahedral coordination gives the 1T polytype while the trigonal prismatic coordination can give the 2H, 3R or 4Hc polytypes. A mixture of the two coordinations can give rise to the 4Hb or the 6R polytypes\cite{Wilson1969,Wilson1974,Kertesz1984}. As a result of their varied crystal structure, often for the same compound, the TMDCs show a rich variety in their electrical and magnetic properties. For example, the Group IV chalcogenide HfS$_2$ is an insulator, most of the Group V dichalcogenides undergo a superconducting transition and are \textit{d} band metals and band antiferromagnets, the Group VI dichalcogenides which have the trigonal prismatic structure are diamagnetic and semiconductors while those with distorted octahedral coordination like WTe$_2$ and $\beta$-MoTe$_2$ are semi--metals\cite{Wilson1969}. In addition, these materials also show charge-density wave (CDW) transitions which has drawn a lot of attention\cite{Bayard1976,Eaglesham1986359,Giambattista1990,Wilson1974}. The layered structure of the TMDCs also leads to quasi two-dimensional nature and marked anisotropy in their electrical properties, along and perpendicular to the layers\cite{Reshak2004,Wilson1974,FLi2014}. Following the success of graphene, the quest for alternative two-dimensional materials has led to renewed interest in the TMDCs because of their quasi two--dimensional nature and because they offer added advantages over graphene as two--dimensional materials in that their properties change when they are monolayers or a few layers thick and these properties can be tuned to suit various requirements\cite{Chhowalla2013,MXu2013,FLi2014,YZhang2016}. Among the TMDCs, samples of MoS$_2$ which are only few layers thick have received considerable attention because of their electrical properties\cite{FLi2014,MXu2013}. More recently, some of the TMDCs, like WTe$_2$ and $\beta$--MoTe$_2$ have been investigated as possible Weyl semi-metals\cite{Soluyanov2015}.

VSe$_2$ is a Group V TMDC that crystallizes in the 1T polytype with V atoms located inside an octahedra of Se atoms\cite{Tsutsumi1982,Thompson1979,Bayard1976,Vanbruggen1976}. VSe$_2$ has a CDW transition at around 110~K which is seen in the temperature dependent electrical transport, Hall, and magnetic susceptibility measurements\cite{Vanbruggen1976,Bayard1976,Thompson1979,DiSalvo1981,Toriumi1981}. Some reports propose that the transition at 110~K is from a normal to an incommensurate CDW state followed by a second transition from an incommensurate to a commensurate CDW state at around 70~K\cite{Thompson1979,Thompson1978}; however, other reports do not find evidence of the transition at the lower temperature\cite{DiSalvo1981,Toriumi1981}. The magnetic susceptibility shows a Curie-Weiss like behaviour at low temperatures which is supposed to originate because of an excess of V ions sitting in the van der Waal spaces between the VSe$_2$ layers in the metal rich VSe$_2$\cite{Thompson1979,Thompson1978,DiSalvo1981,Vanbruggen1976,Bayard1976}. It is suggested that each interstitial V ion produces a net paramagnetic moment of 2.5 Bohr magnetons and the atomic percentage of the interstitial V ions can be determined using this value and the value of the Curie constant\cite{DiSalvo1981,Thompson1979,Thompson1978}. According to DiSalvo and Waszczak the CDW transition temperature is suppressed at a rate of $13\pm3$~K/at $\%$ of V interstitial ions\cite{DiSalvo1981}. The CDW transition is also reported to be suppressed with decreasing thickness in nanoflakes of VSe$_2$\cite{Yang2014}. The resistivity of VSe$_2$ at low temperatures has variously been reported to show both $T^2$ and $T^4$ dependence\cite{Bayard1976,Thompson1979,Toriumi1981}. The Hall constant is $n$-type and shows a large increase with temperature below the CDW transition temperature\cite{Toriumi1981,Bayard1976,Thompson1979,Vanbruggen1976}. The transverse magnetoresistance in VSe$_2$, in fields either perpendicular or parallel to the layers of the sample shows a monotonic rise with field and does not saturate\cite{Toriumi1981}.

At low temperatures, quantum effects sometimes become visible in the macroscopic properties such as the resistivity of a material. One such example is the low-temperature upturn in resistivity which is seen in a variety of systems and can originate from quantum corrections to resistivity due to weak localization or electron-electron interactions\cite{Gao2012,Niu2016,Tiwari1999,Kumar2002,Maritato2006,Xu2006}. A similar upturn in resistivity at low temperatures is also seen in the Kondo effect which is due to the scattering of conduction electrons by localized magnetic impurities through the $s$-$d$ exchange interaction\cite{Xu2006,Katayama1967}. In this paper, we report a low-temperature upturn in the resistivity of single crystal samples of VSe$_2$ and a negative magnetoresistance that is seen at low fields and temperatures below the minimum in resistivity. The VSe$_2$ single crystal samples were characterized using single crystal X-ray diffraction, Laue X-ray diffraction, resistivity, Hall effect, DC magnetization measurements and specific heat measurements. The low-temperature rise in resistivity and the negative magnetoresistance seen at low fields and temperature are analysed in conjunction with the magnetization and heat capacity data to determine if they are due to the Kondo effect arising from a scattering of the conduction electrons by the paramagnetic V ions present in the van der Waal spaces in between the layers of VSe$_2$.

\section*{Results}

Polycrystalline powder samples of VSe$_2$ as well as single crystals of VSe$_2$ were synthesized. First, the results of the polycrystalline VSe$_2$ are presented, followed by those for the single crystals of VSe$_2$.

\begin{figure}[tb]
\centering
\includegraphics[width=\linewidth]{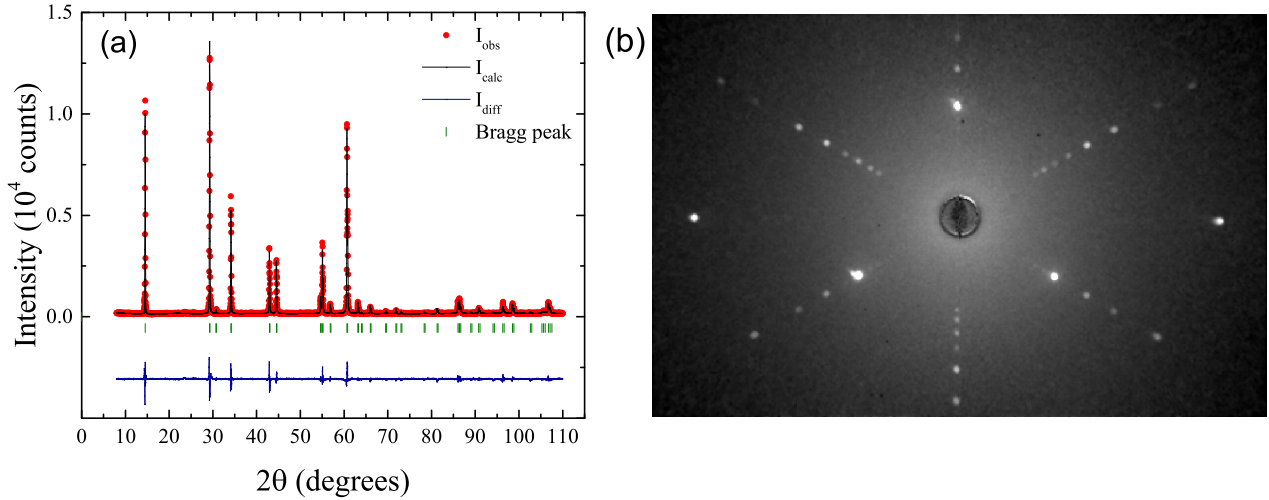}
\caption{(a) Room temperature powder X-ray diffraction pattern of VSe$_{2}$. The experimental profile (red closed circles) and a full profile matching refinement (black solid line) made using the $P\bar{3}m1$ hexagonal structure are shown, with the difference given by the blue solid line. (b) X-ray Laue back scattered diffraction pattern for a single crystal of VSe$_2$ viewed along the [001] direction.}
\label{fig:Polycrystalline_XRD_n_Laue}
\end{figure}

\subsection*{Polycrystalline VSe$_2$}

\begin{figure}[ht]
\centering
\includegraphics[width=\linewidth]{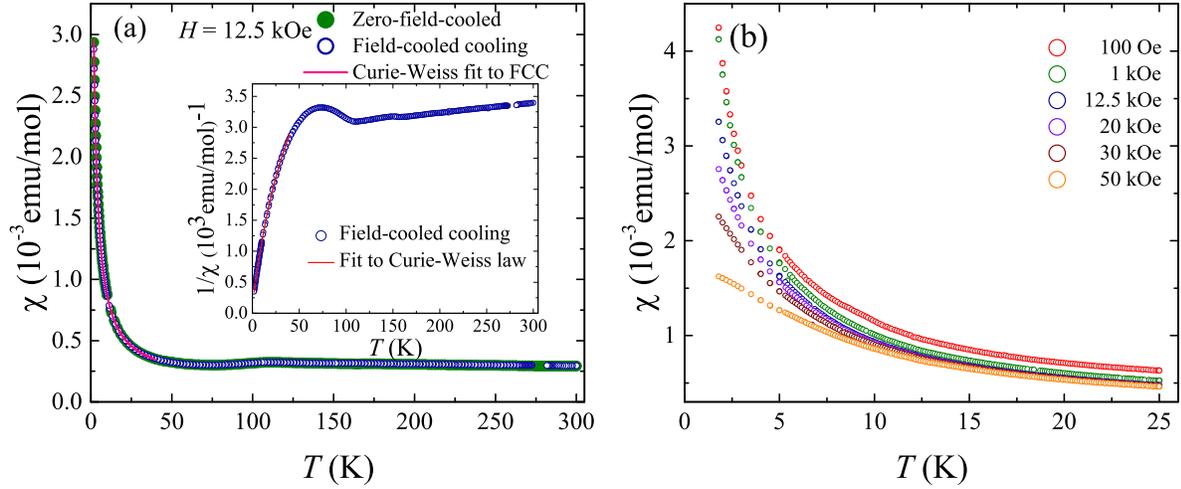}
\caption{(a) Susceptibility versus temperature and its inverse (inset) for a polycrystalline sample of VSe$_2$. The fit to a modified Curie-Weiss law from 1.8--40~K for both the susceptibility versus temperature and its inverse are also shown. (b) Low-temperature susceptibility for the polycrystalline sample of VSe$_2$ at different applied magnetic fields.}
\label{fig:Polycrystalline_Chi_vs_T}
\end{figure}

Room temperature powder X-ray diffraction measurements were performed to determine the phase purity of the polycrystalline material. The X-ray data were fitted to the $P\bar{3}m1$ (No. 164) space group using the FullProf software suite\cite{Rodriguez1993}. The X-ray diffraction profile for the VSe$_{2}$ polycrystalline material displayed in Fig.~\ref{fig:Polycrystalline_XRD_n_Laue}(a) shows no trace of any impurity and reveals that the hexagonal structure of the 1T polytype is formed\cite{Bayard1976}. The relative intensities of some of the Bragg peaks ($hkl$ = (001), (002), (003) and (004)) are significantly increased for polycrystalline samples, due to the preferred orientation present in the sample. The lattice parameters ($a = 3.35844(3)$~\r{A} and $c = 6.10141(5)$~\r{A}) were found to be similar to the published values for VSe$_2$\cite{Bayard1976,Wilson1969}.

Figure~\ref{fig:Polycrystalline_Chi_vs_T}(a) shows the susceptibility and the inverse susceptibility versus temperature for the polycrystalline VSe$_2$ sample. The data reveal a prominent upturn at low temperatures and a change in slope around 112~K corresponding to the CDW transition. We have fitted both the susceptibility data and its inverse from 1.8 - 40 K to a modified Curie-Weiss law
\begin{equation}
\chi\left(T\right) = \frac{\mathrm{C}}{T-\Theta}+\chi_0,
\label{Curie_Weiss_law}
\end{equation}
which consists of a temperature independent term in addition to the Curie-Weiss part.
The results of the fit are shown in the inset of Fig.~\ref{fig:Polycrystalline_Chi_vs_T}(a). From the fit to the inverse susceptibility data, we obtain the values of {$\mathrm{C} = (0.00797 \pm 0.00003)$~emu--K/mol, $\Theta = (-1.04\pm0.02)$~K and $\chi_0=(1.51\pm 0.01) \times 10^{-4}$~emu/mol. We discuss this in more detail later, in the section on the susceptibility versus temperature data for the single crystal sample. Figure~\ref{fig:Polycrystalline_Chi_vs_T}(b) shows the low-temperature susceptibility for different applied magnetic fields for the polycrystalline sample. The magnetic susceptibility decreases with increasing applied field and there is no clear signature of any magnetic order down to the lowest temperatures measured.

\subsection*{Single crystals of VSe$_2$}

Single crystals of VSe$_2$ were studied using electrical transport, Hall, magnetoresistance, DC magnetization, and heat capacity measurements. In Fig.~\ref{fig:Polycrystalline_XRD_n_Laue}(b) we show the X-ray Laue back scattered diffraction pattern of a single crystal of VSe$_2$ obtained along the [001] direction. The pattern shows the good quality of the single crystal. In order to confirm the crystal structure of our VSe$_2$ samples, single crystal X-ray diffraction measurements were performed on the single crystals. A refinement of the single crystal X-ray diffraction data was performed using the space group $P\bar{3}m1$ (No. 164) and a goodness of fit (GooF) of 1.068 was obtained. The values of the crystal parameters obtained were $a = b = 3.357(1)$~\r{A}, $c = 6.104(5)$~\r{A} and a cell volume of 59.59(6)~\r{A}$^3$ which are in agreement with previously reported results\cite{Wilson1969}. We also performed compositional analysis on the single crystals using energy dispersive X-ray spectroscopy (EDS) which yielded an average composition of $\mathrm{V} = \left(28\pm 2\right) \%$ and Se$ = \left(72\pm 2\right) \%$, showing a V deficiency in the overall stoichiometry. Below, we discuss the results of our measurements on single crystals of VSe$_2$.

\begin{figure}[ht]
\centering
\includegraphics[width=\linewidth]{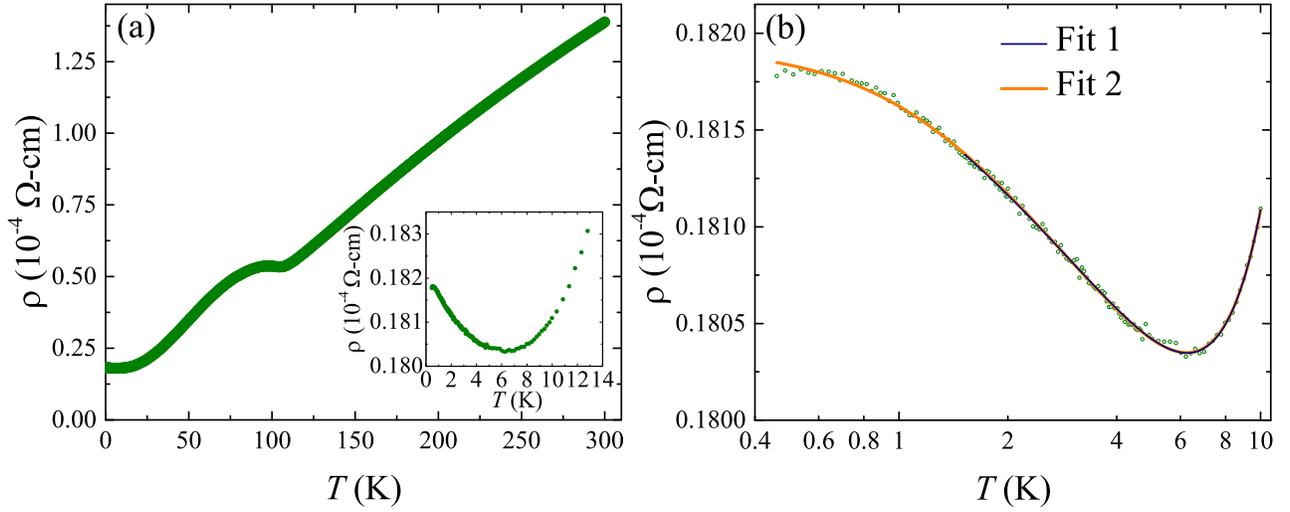}
\caption{(a) Resistivity versus temperature for a single crystal of VSe$_2$. The inset shows the low-temperature upturn in resistivity. (b) The low temperature upturn in resistivity plotted versus $\ln T$. Fit 1 is the result of fit in the temperature range 1.5 to 10~K using Equation~\ref{Equation_for_Fit}. Fit 2 is the result of fitting over the temperature range 0.45 to 10~K and replacing $T$ with $T_{\mathrm{eff}}$ in the expression for $\rho_{\mathrm{sd}}$ in Equation \ref{Equation_for_Fit}. $T_{\mathrm{K}}$ and $S$ are kept fixed at the values obtained in Fit 1.}
\label{fig:RvsT}
\end{figure}

\subsubsection*{Low-temperature upturn in resistivity}

Figure~\ref{fig:RvsT}(a) shows the variation of resistivity with temperature, $\rho\left(T\right)$, down to 0.45~K for a single crystal of VSe$_2$. It is evident that the single crystal has a metallic nature with a plateau around 110~K which marks the CDW transition in VSe$_2$. However, instead of flattening off to a residual value at low temperatures, as is expected for a simple metal, the resistivity exhibits a minimum at around 6.5~K and then increases as the temperature decreases until about 0.5~K where $\rho\left(T\right)$ shows signs of saturation (see the inset of Fig.~\ref{fig:RvsT}(a)). We have measured the resistivity versus temperature of crystals taken from two separate growths and they all show an upturn in resistivity at low temperature. We show some of these results in the supplementary information. Such a low-temperature minimum in $\rho\left(T\right)$ can occur due to electron-electron interactions, weak localization, or the Kondo effect\cite{Niu2016,Barone2015,Zhang2009,Tiwari1999}.

For three dimensional systems, an increase in resistivity with decreasing temperature should follow either a $T^{-p/2}$ dependence if it is due to weak localization ($p = \frac{3}{2}$, 2, 3 depending upon the scattering mechanism)\cite{Lee1985,Maritato2006}, a $-\sqrt{T}$ dependence if it is due to electron-electron interactions \cite{Maritato2006}, or a $-\ln T$ dependence if it is due to the Kondo effect\cite{Kumar2002,Xu2006}. Weak localization and electron-electron interactions can be ruled out as possible reasons for the observed upturn in resistivity as fits to a $-T^{3/4}$ (weak localization in the dirty limit of dominant electron-electron scattering) and a $-T^{1/2}$ behaviour produced poor agreement with the experimental data.

Figure~\ref{fig:RvsT}(b) shows the low-temperature upturn in resistivity versus $\ln T$, from which it is clear that for a certain range of temperature the upturn in resistivity is linear in $\ln T$, as expected for the Kondo effect, and only begins to flatten off at the lowest temperatures measured. The deviation from a $-\ln T$ dependence may be the result of the formation of a Nagaoka spin-compensated state or Ruderman-Kittel-Kasuya-Yosida (RKKY) mediated interactions between the magnetic impurities responsible for the Kondo effect~\cite{Kastner1977,Liang1973}.

We carefully examined the fitting of the temperature dependence of resistivity within the framework of the Kondo effect. The Kondo resistivity in the region where $\rho\left(T\right)$ is linear in $\ln T$ can be described by the Hamann expression~\cite{Hamann1967,Liang1973},

\begin{equation}
\rho_{sd} = \frac{\rho_{\mathrm{0}}}{2}\left(1-\frac{\ln\left(T/T_{\mathrm{K}}\right)}{\left[\ln^2\left(T/T_{\mathrm{K}}\right)+S(S+1)\pi^2\right]^{1/2}}\right)
\label{Equation_Hamann_equation}
\end{equation}
\noindent where, $\rho_{\mathrm{0}}$ is the unitarity limit, $T_{\mathrm{K}}$ is the Kondo temperature and $S$ is the spin of the magnetic impurity. We account for any remaining temperature dependence in $\rho\left(T\right)$ due to electron-phonon scattering at temperatures below the resistivity minimum, by first fitting the $\rho\left(T\right)$ data in the temperature range 10 to 20~K to an expression of the form $a + b T^n$. We obtain the numerical values for $b$ and $n$ equal to $(6.8\pm0.7)\times 10^{-11}$ and $3.37\pm0.03$, respectively. The low-temperature resistivity is then fit over the range 1.5 to 10~K using an expression that includes the Hamann term, along with an electron-phonon contribution $b T^n$ and a temperature independent term $\rho_b$,
\begin{equation}
\rho(T) = \rho_{sd} + bT^n + \rho_{\mathrm{b}}.
\label{Equation_for_Fit}
\end{equation}
While performing this fit, the coefficient $b$ and power $n$ of the electron-phonon contribution are held fixed at the values obtained from the higher temperature $\rho\left(T\right)$ data. The result is shown in Fig. \ref{fig:RvsT}(b) as Fit 1 and the parameters obtained are listed in Table~\ref{Table1}.

The Hamann expression is known to give a poor fit for temperatures $T < T_{\mathrm{K}}$, where the resistivity deviates from a $\ln T$ behaviour and can give low values for $S$ as a result of inadequacies in the Nagaoka approximation on which the Hamann expression is based~\cite{Liang1973,Apostolopoulos1996}. The deviation of the resistivity from the $\ln T$ behaviour at low temperature can be accounted for by considering the RKKY interactions between the paramagnetic interlayer V ions. The low-temperature resistivity can be fit to a modified Equation~\ref{Equation_for_Fit} where the temperature $T$ in the expression for $\rho_{\mathrm{sd}}$ is replaced by an effective temperature $T_{\mathrm{eff}} = \left(T^2+T_{\mathrm{W}}^2\right)^{1/2}$ and $k_{\mathrm{B}}T_{\mathrm{W}}$ is the effective RKKY interaction strength~\cite{Kastner1977}. For this second fit, $S$ and $T_{\mathrm{K}}$ along with $b$ and $n$ were fixed to the values obtained in Fit 1. The result of Fit 2 is also shown in Figure~\ref{fig:RvsT}(b) and provides a satisfactory description of the data across the entire temperature range 0.45 to 10~K. The parameters obtained from Fit 2 are listed in Table~\ref{Table2}. The value of 1.18~K obtained for the temperature $T_\mathrm{W}$  is consistent with the temperature at which the resistivity starts deviating from a $\ln T$ behaviour.

\begin{table}[ht]
\centering
\begin{tabular}{|l|l|l|l|}
\hline
$\rho_{\mathrm{0}}$ ($\Omega$-cm) & $\rho_{\mathrm{b}}$ ($\Omega$-cm) & $T_{\mathrm{K}}$ (K) & $S$  \\
\hline
$(5.9\pm0.7)\times10^{-7}$ & $(1.770\pm0.005)\times10^{-5}$ & $6.8\pm0.9$ & $0.48\pm0.09$ \\
\hline
\end{tabular}
\caption{\label{Table1} Parameters obtained from a fit to the resistivity versus temperature using Equation~\ref{Equation_for_Fit} in the temperature range 1.5 to 10~K.}
\end{table}

\begin{table}[ht]
\centering
\begin{tabular}{|l|l|l|}
\hline
$\rho_{\mathrm{0}}$ ($\Omega$-cm) & $\rho_{\mathrm{b}}$ ($\Omega$-cm) & $T_{\mathrm{W}}$ (K) \\
\hline
$(6.08\pm0.03)\times10^{-7}$ & $(1.7691\pm0.0002)\times10^{-5}$ & $1.18\pm0.02$ \\
\hline
\end{tabular}
\caption{\label{Table2} Parameters obtained from the fit of the resistivity versus temperature between 0.45 and 10~K using Equation~\ref{Equation_for_Fit} with $T$ replaced by effective temperature $T_{\mathrm{eff}}$ in the expression for $\rho_{\mathrm{sd}}$ in order to take into account the effect of RKKY interaction between the interlayer V ions.}
\end{table}

\begin{figure}[ht]
\centering
\includegraphics[width=\linewidth]{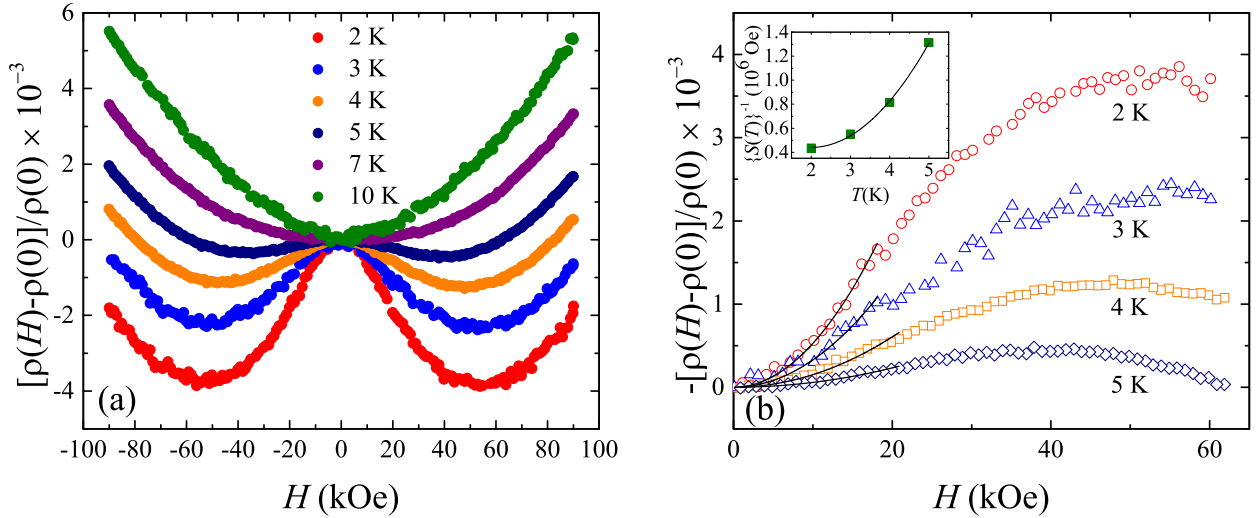}
\caption{(a) Relative magnetoresistance at different temperatures. A negative magnetoresistance is seen at low fields and low temperatures but disappears as the temperature increases. (b) Fit of the low field negative magnetoresistance at low temperatures to a parabolic dependence on magnetic field. The inset shows the variation of the parameter $\left(S\left(T\right)\right)^{-1}$ obtained from the fit showing a parabolic dependence for $(S(T))^{-1}$ with temperature.}
\label{fig:RvsH}
\end{figure}

\subsubsection*{Negative magnetoresistance}
The magnetoresistance of the samples was measured with the magnetic field parallel to the $c$-axis of the single crystal samples. Figure~\ref{fig:RvsH}(a) shows the relative magnetoresistance at different temperatures. It can be seen that the magnetoresistance below 7~K is negative at low fields but starts showing a positive trend as the field increases further. At 7~K and above, the magnetoresistance is positive for all fields. Thus the negative magnetoresistance seen at low fields occurs in the same temperature range at which a rise in resistivity is seen with falling temperature. This suggests that the negative magnetoresistance and the low-temperature upturn in resistivity are correlated and could arise due to the Kondo effect.

We further analyse the negative magnetoresistance to see if it could be due to the Kondo effect. Negative magnetoresistance arising due to the Kondo effect is reported to show a parabolic dependence with the magnetic field at low fields\cite{Katayama1967},
\begin{equation}
-\Delta \rho / \rho = -\left[\rho\left(H\right) - \rho\left(0\right)\right]/\rho\left(0\right) = S\left(T\right) H^2,
\label{Equation_Kondo_MR}
\end{equation}
where $S(T)$ is the coefficient of the magnetoresistance. In Fig.~\ref{fig:RvsH}(b), we show the fit to the relative negative magnetoresistance up to a field of 20~kOe at different temperatures using Eq.~\ref{Equation_Kondo_MR}. In the case of InSb, a similar negative magnetoresistance was seen due to the Kondo effect and the coefficient $S\left(T\right)$ in that case increased with temperature and varied as $\left(S\left(T\right)\right)^{-1/2} \propto \left(T+T_0\right)$ where $T_0$ is a parameter\cite{Katayama1967}. However, for our samples, $\left(S\left(T\right)\right)^{-1}$ shows a parabolic temperature dependence as shown in the inset of Fig.~\ref{fig:RvsH}(b).

\begin{figure}[tb]
\centering
\includegraphics[width=\linewidth]{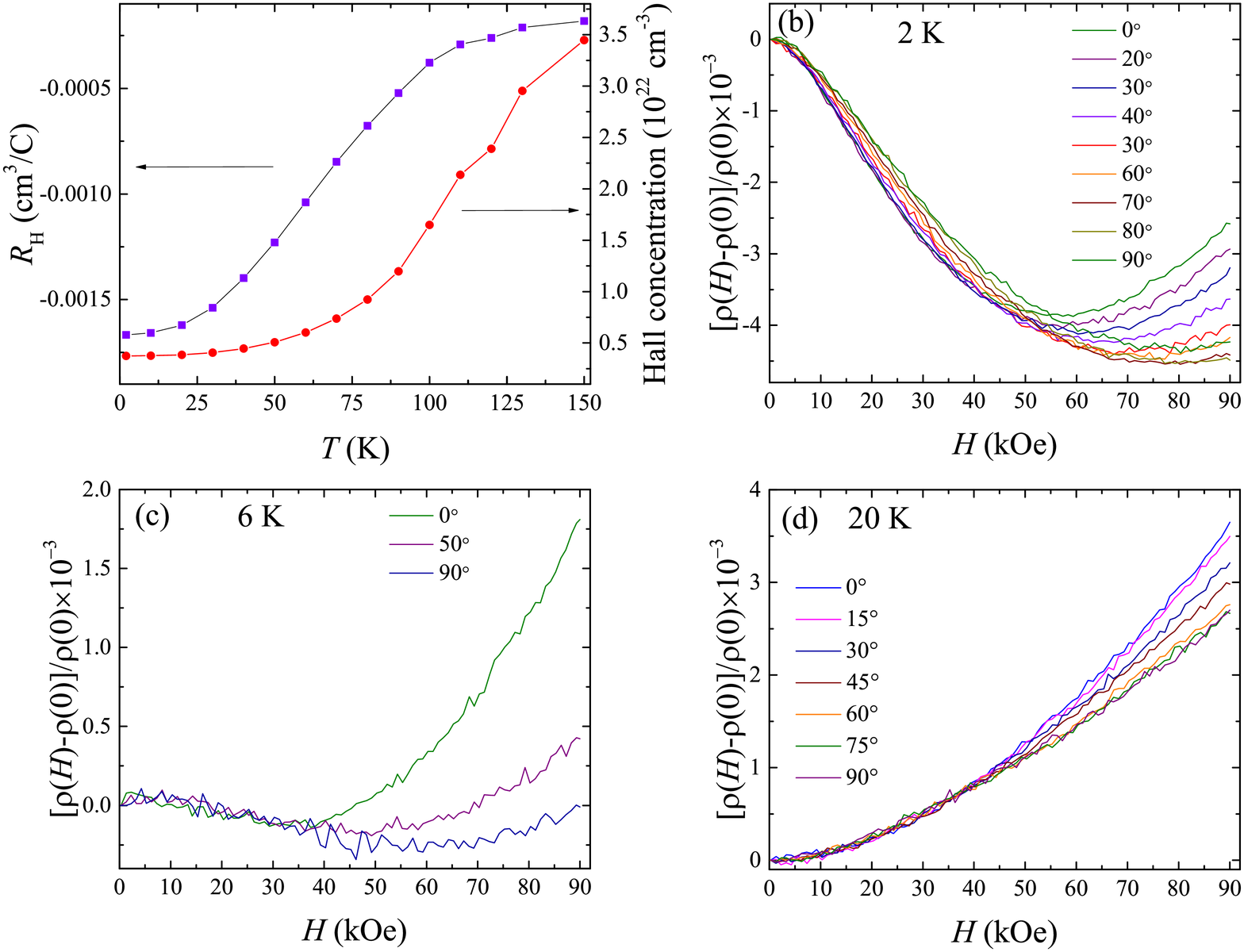}
\caption{(a) Variation of Hall coefficient and Hall concentration with temperature. The dependence of the relative change in transverse magnetoresistance at (b) 2~K , (c) 6~K and (d) 20~K on the angle between the field and the perpendicular to the plane of the sample. The raw data were symmetrized to remove a contribution from the Hall effect.}
\label{fig:Hall_vs_T_n_Angle_Dependence_MR}
\end{figure}

\subsubsection*{Hall concentration versus temperature}
Figure~\ref{fig:Hall_vs_T_n_Angle_Dependence_MR}(a) shows the variation  with temperature of the Hall constant, $R_\mathrm{H}$, and the Hall carrier concentration, $n$, obtained using the relation $R_{\mathrm{H}}= -1/ne$. The Hall constant is negative, indicating that the sample is $n$-type. At temperatures immediately below the CDW transition seen in the resistivity versus temperature, the Hall constant shows a considerable increase in magnitude and the Hall concentration, a significant decrease. VSe$_2$ has been reported to have $n$-type charge carriers and a similar increase in the magnitude of the Hall constant at the CDW transition\cite{Vanbruggen1976,Thompson1979,Toriumi1981}. A CDW transition leads to a decrease in the density of states at the Fermi level which in turn reduces the carrier concentration\cite{Vanbruggen1976}. The magnitude of the Hall constant in our sample at low temperature is ${\left(1.7\pm 0.3\right)\times 10^{-3}~\mathrm{cm}^3/\mathrm{C}}$, while the values of between $1.1\times 10^{-3}$ and ${4.2\times 10^{-3}~\mathrm{cm}^3/\mathrm{C}}$ are reported in this material\cite{Vanbruggen1976,Thompson1979,Bayard1976,Toriumi1981,Yadav2010}.

\subsubsection*{Angular dependence of magnetoresistance}
We also measured the angular dependence of the transverse magnetoresistance at different temperatures which is shown as relative change in magnetoresistance at different angles in Figs.~\ref{fig:Hall_vs_T_n_Angle_Dependence_MR}(b), \ref{fig:Hall_vs_T_n_Angle_Dependence_MR}(c), and \ref{fig:Hall_vs_T_n_Angle_Dependence_MR}(d) for the temperatures 2, 6, and 20~K respectively. VSe$_2$ being a layered material cleaves perpendicular to the $c$-axis and hence in our samples, the $c$-axis is perpendicular to the plane of the sample. The magnetic field is rotated in a plane containing the $c$-axis and the direction perpendicular to the current. The angular dependence of the magnetoresistance at 2 and 6~K are shown in Figs.~\ref{fig:Hall_vs_T_n_Angle_Dependence_MR}(b) and \ref{fig:Hall_vs_T_n_Angle_Dependence_MR}(c). From the figures, it is evident that the negative magnetoresistance seen at low fields does not vary much with the change in the angle the field makes with the $c$-axis of the sample. However, the positive trend seen in the magnetoresistance at higher fields decreases significantly in magnitude as the field becomes perpendicular to the $c$-axis. At the higher temperature of 20~K, (see Fig.~\ref{fig:Hall_vs_T_n_Angle_Dependence_MR}(d)), where the negative magnetoresistance has already disappeared, the positive magnetoresistance decreases in magnitude as the field becomes perpendicular to the $c$-axis. This kind of angular dependence of the transverse magnetoresistance is another indication that the low-temperature upturn in resistivity is due to the Kondo effect with an isotropic $s$--$d$ exchange interaction between the conduction electrons and the localized paramagnetic spins because in such a case the negative magnetoresistance does not show much dependence on the direction of the field\cite{Katayama1967,Zhang2009}. Toriumi and Tanaka\cite{Toriumi1981} reported that the magnetoresistance in VSe$_2$ at low fields was smaller for fields parallel to the $c$-axis than for fields perpendicular to the $c$-axis and vice versa at higher fields. However, we do not observe this effect in our sample.

\subsubsection*{Magnetization measurements}
\label{Magnetization measurement}

\begin{figure}[tb]
\centering
\includegraphics[width=\linewidth]{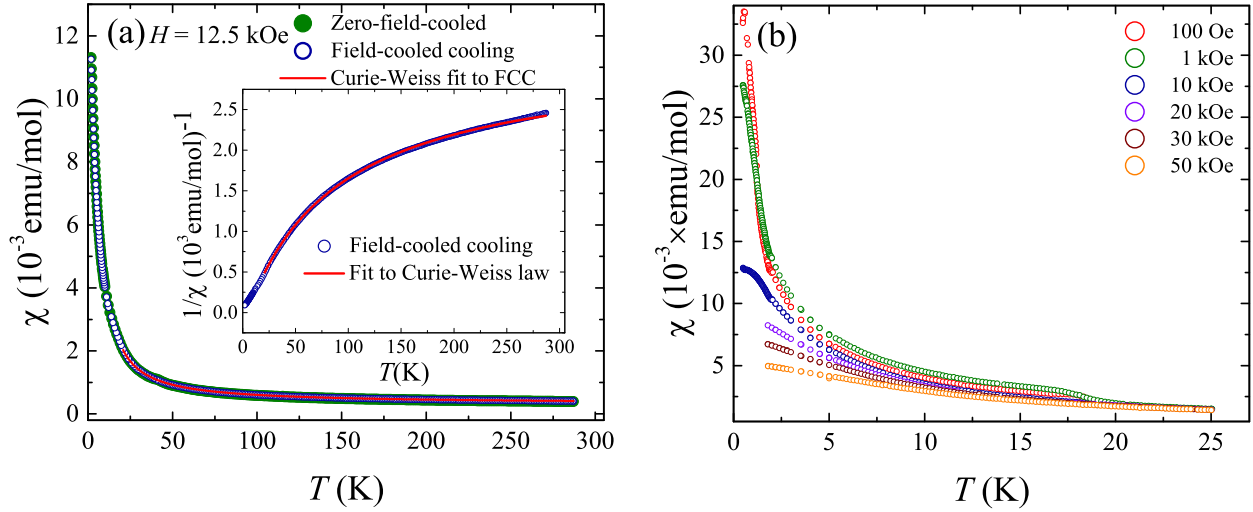}
\caption{(a) Susceptibility versus temperature for a collection of single crystal samples of VSe$_2$ with a magnetic field applied perpendicular to the $c$-axis of the crystals. The inset shows the inverse susceptibility versus temperature. Fits to a Curie-Weiss law along with a temperature independent term for both the field-cooled susceptibility and inverse susceptibility data above 20~K are also shown. (b) Low-temperature susceptibility at different applied magnetic fields for single crystals of VSe$_2$ which were co-aligned with their $c$-axis perpendicular to the magnetic field. The measurements at 100~Oe, 1~kOe, and 10~kOe in the temperature range 0.5 to 2~K were performed separately and the data were adjusted to remove the background signal from the \textit{i}Quantum $^{3}$He insert.}
\label{fig:CurieWeiss_fit_crystal}
\end{figure}

Figure~\ref{fig:CurieWeiss_fit_crystal}(a) shows the results of the susceptibility versus temperature measurement for several single crystals of VSe$_2$, co-aligned with their $c$-axis pointing perpendicular to the magnetic field direction. Unlike the polycrystalline data, where a change in slope is clearly seen around 112~K at the CDW transition, there is no such prominent transition in the susceptibility data for the single crystals. However, the susceptibility shows an upturn as temperature decreases similar to that seen in the case of the polycrystalline powder. Such an upturn in the susceptibility in VSe$_2$ has previously been attributed to the presence of V ions in the van der Waal gaps which possess a net paramagnetic moment\cite{DiSalvo1981}. We analysed the susceptibility and the inverse susceptibility data for the single crystal samples by fitting it to the modified Curie-Weiss law given in Eq.~\ref{Curie_Weiss_law}. Figure~\ref{fig:CurieWeiss_fit_crystal}(a) shows the results of these fits for both the field-cooled susceptibility and inverse susceptibility data for the single crystals. We fit only the data above 20~K, as the data clearly deviate from a  Curie-Weiss behaviour at lower temperatures (see below). From the fit to the inverse susceptibility data, we obtain the values of $\mathrm{C}=(0.02773\pm 0.00007)$~emu--K/mol, $\Theta = (4.59 \pm 0.06)$~K and $\chi_0=(3.143\pm 0.004)\times 10^{-4}$~emu/mol. The temperature independent term accounts for a major part of the susceptibility at high temperatures whereas at lower temperatures the Curie-Weiss term plays an increasingly important role. It has been suggested that the V ions in the interlayer gaps each possess a net paramagnetic moment of 2.5~Bohr magnetons. Using this, the molar fraction of the V ions in the interlayer gaps can be calculated. Using the value of the Curie constant obtained from the fit to the susceptibility versus temperature, we arrive at the molar fraction of interlayer V ions of $0.0102\pm 0.0004$ and $0.0357\pm 0.0001$, respectively, for the polycrystalline and the single crystal samples.

In Fig.~\ref{fig:CurieWeiss_fit_crystal}(b), we show the low-temperature susceptibility at different fields for single crystal VSe$_2$, with the data at 100~Oe, 1~kOe, and 10~kOe measured down up to 0.5~K. In the lower fields the susceptibility contains a clear anomaly at around 20~K that is not evident in the higher field curves. This feature at 20~K is not present in the $\chi\left(T\right)$ data for the polycrystalline samples that contain a lower concentration of interlayer V ions as is evident from Fig.~\ref{fig:Polycrystalline_Chi_vs_T}(b). The measurements at 100~Oe, 1~kOe, and 10~kOe, and in particular the one at 10~kOe, show that the susceptibility deviates significantly from a Curie-Weiss law at low temperatures. The 100~Oe measurement shows a maximum at 0.5~K.  The step at 20~K, the deviations from Curie-Weiss behaviour, as well as the low-temperature maximum in $\chi\left(T\right)$ are the result of the onset of magnetic correlations, short-range magnetic order, or clustering between the V ions, mediated by an RKKY interaction between the interlayer V ions, combined with a reduction in the effective moment of the V ions due to Kondo screening by the conduction electrons. These magnetic correlations and the reduction in the Kondo screened moments are also reflected in the deviation from $\ln T$ behaviour seen in the resistivity.

\subsubsection*{Specific heat measurements}

\begin{figure}[t]
\centering
\includegraphics[width=\linewidth]{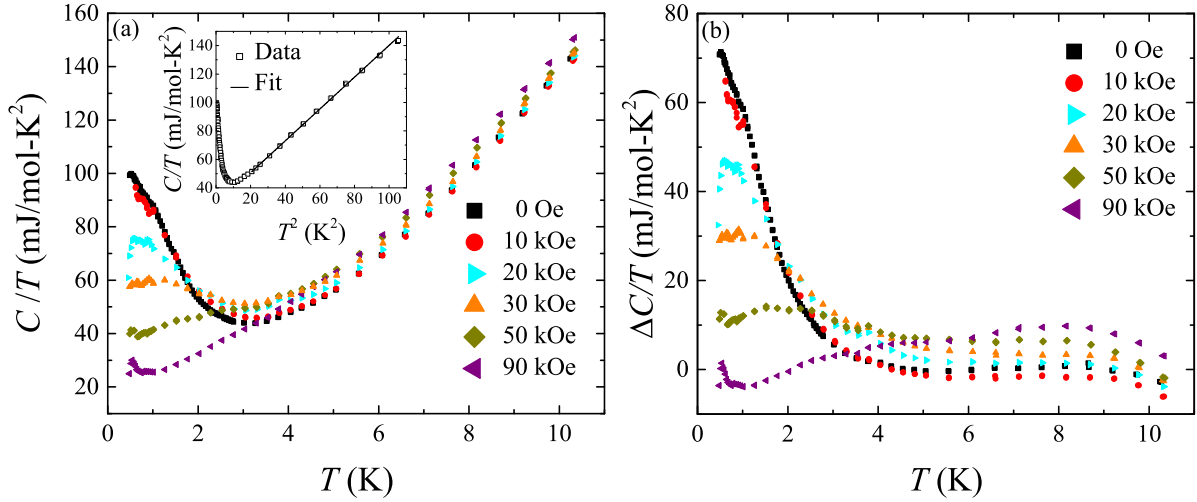}
\caption{(a) Specific heat capacity divided by temperature versus temperature at different magnetic fields for several single crystals of VSe$_2$ co-aligned together with their $c$-axis pointing perpendicular to the applied magnetic field. Inset: Plot of $C/T$ versus $T^2$ at zero field and fit to expression $C/T = \gamma + \beta T^2$. (b) Excess specific heat divided by temperature $\Delta C/T$ versus temperature obtained by subtracting the electronic and lattice contributions to specific heat in different magnetic fields.}
\label{fig:CbyTvsT_DeltaCbyTvsT}
\end{figure}

We measured the specific heat of our samples to shed more light on their magnetic and electronic properties. The specific heat of several crystals of VSe$_2$, which were co-aligned so that the magnetic field was perpendicular to the $c$-axis of the crystals, was measured from 0.5 to 10~K in different applied magnetic fields. In Fig.~\ref{fig:CbyTvsT_DeltaCbyTvsT}(a), we show the results of these measurements, plotting the specific heat divided by temperature versus temperature at different magnetic fields. At zero magnetic field, the specific heat divided by temperature shows a rise at low temperatures. Such a low-temperature rise in the specific heat can be due to a variety of phenomena such as the Schottky anomaly due to transitions between levels in a system of paramagnetic moments which are either intrinsic or present as impurities\cite{Gopal2012,Cava1993}, Kondo behaviour\cite{Yadam2016,Anand2015}, magnetic clusters\cite{Aitken1981}, and spin fluctuations\cite{Dhar1987}. Generally in Kondo systems, the rise in specific heat at low temperatures given by the $\gamma$ value obtained by extrapolating $C/T$ vs. $T^2$ to $T = 0$ depends on the number of Kondo scattering centres and is inversely proportional to the Kondo temperature $T_{\mathrm{K}}$.  $\gamma$ values of between a few mJ/mol-K$^2$  and 2500~mJ/mol-K$^2$ are not uncommon in Kondo systems\cite{Yadam2016,Hillier2007,Jaime2000}. This value of $\gamma$ for our samples at zero field is $102\pm2$~mJ/mol-K$^2$. The high field value of $\gamma$ of $\approx$ $32\pm2$~mJ/mol-K$^2$ is also large, indicating the presence of Kondo scattering. The application of a magnetic field suppresses the low-temperature rise in specific heat in our samples and shifts the anomaly to higher temperatures, as can be seen in Fig.~\ref{fig:CbyTvsT_DeltaCbyTvsT}(a). This can be analysed better by subtracting the electronic and lattice contributions to specific heat from the total specific heat. To do this, we first plot the zero field $C/T$ versus $T^2$ and fit the high temperature part to the expression $C/T = \gamma + \beta T^2$, where $\gamma T$ is the high field electronic contribution and $\beta T^3$ the lattice contribution to the specific heat. This fit is shown in the inset of Fig.~\ref{fig:CbyTvsT_DeltaCbyTvsT}(a). We obtain a value of $\gamma=28.3\pm0.2$~mJ/mol-K$^2$ for our sample which is still larger than the values of between 3 and 16~mJ/mol-K$^2$ reported for other TMDCs\cite{Harper1977}. The Debye temperature $\theta_{\mathrm{D}}$ can be calculated from the $\beta$ value using the expression $\beta = 12nR\pi^4/5\theta_{\mathrm{D}}^3$, where $R$ is the universal gas constant and $n$ the number of atoms per molecule. The $\beta$ value of 1.119~$\pm$~0.005~mJ/mol-K$^4$ for our sample gives a Debye temperature, $\theta_{\mathrm{D}}$ of $173.3\pm0.3$~K which is comparable to that of 213~K reported for VSe$_2$\cite{Yadav2010}. The values of the electronic and lattice contribution to specific heat obtained from the fit are then subtracted from the total specific heat measured at different fields including zero field, assuming that the $\gamma$ and $\beta$ values do not change with field. This gives the excess specific heat $\Delta C$ which is divided by temperature and plotted against temperature as shown in Fig.~\ref{fig:CbyTvsT_DeltaCbyTvsT}(b). From the figure, it is clear that the magnitude of the maximum in the excess specific heat at low temperatures goes down as the field is increased and that the peak shifts to higher temperatures as the field increases. This kind of behaviour of the excess specific heat is consistent with the evidence for RKKY mediated magnetic interactions seen in both the low-field low-temperature $\chi\left(T\right)$ data shown in Fig.~\ref{fig:CurieWeiss_fit_crystal}(b) and the low-temperature $\rho\left(T\right)$ curve presented in Fig.\ref{fig:RvsT}(b). The area under the $\Delta C/T$ versus $T$ gives the entropy associated with this Schottky-like anomaly and for our sample at 30~kOe this is $92\pm2$~mJ/mol-K which amounts to 0.010 of the entropy $R\ln(2S +1)$ associated with a mole of paramagnetic spins with spin $S = 1$, where $R$ is the molar gas constant (8.314~J/mol-K). It should be noted that the spin corresponding to the magnetic moment of 2.5 Bohr magneton possessed by the interlayer V ions is $S =0.85$. Bearing in mind the susceptibility results from which we found the molar fraction of interlayer V ions in the single crystals to be 0.0357, and the fact that the method used may underestimate the entropy associated with this anomaly, the heat capacity data are consistent with anomaly arising from magnetic cluster of the interlayer V ions, while an enhanced $\gamma$ even at 90~kOe, is consistent with presence of Kondo scattering.

\section*{Discussion}

The upturn in the low-temperature resistivity in our samples is best described by the Kondo effect and the fit to the low-temperature resistivity using the Hamann equation with an effective temperature that takes into account the RKKY interactions between the magnetic impurities gives a convincing description of the $\rho\left(T\right)$ data. The temperature dependence of the resistivity cannot be fit to the behaviour expected in the case of electron-electron interactions or weak localization effects. Moreover, the negative magnetoresistance seen in our sample at low fields and at temperatures below the minimum in the resistivity is further evidence against electron-electron interactions being the source of the upturn in resistivity as the upturn would then be expected to be largely insensitive to magnetic fields\cite{Xu2006,Niu2016} and only exhibit a small positive magnetoresistance\cite{Kumar2002,Maritato2006}. For both the Kondo effect and weak localization, the low-temperature rise in resistivity is expected to disappear in a magnetic field\cite{Kumar2002,Niu2016} and both effects lead to a negative magnetoresistance\cite{Maritato2006,Katayama1967}. However, the negative magnetoresistance in our samples also shows a quadratic dependence with the magnetic field at low fields, similar to what is seen in InSb due to the Kondo effect\cite{Katayama1967}. Furthermore, the magnetoresistance is not sensitive to the angle that the magnetic field makes with the plane of the sample, which is also expected in the case of a Kondo effect with an isotropic $s$--$d$ exchange interaction. These facts taken together suggest that the negative magnetoresistance originates from a Kondo effect. The presence of paramagnetic interlayer V ions indicated by the magnetization and heat capacity measurements means that the localized magnetic moments needed for the scattering of the conduction electrons are present in our samples. Since, the presence of interlayer V ions is necessary for the Kondo effect it implies a certain degree of disorder is required for this effect to be seen. The localised moments appear to suppress the anomaly at the CDW transition seen in the $\chi\left(T\right)$ curves for the polycrystalline sample. A similar behaviour has been observed in the $\chi\left(T\right)$ data for 2$H$-Pd$_x$TaSe$_2$ system, where the small local moments induced by the Pd intercalation lead to a pronounced Curie-Weiss tail in the magnetic susceptibility at low temperature and wash out the transition seen in $\chi\left(T\right)$ at the incommensurate CDW transition around $\sim112$~K in the pure 2H-TaSe$_2$ material\cite{Bhoi2016}. The enhanced $\gamma$, even at 90~kOe, provides further support for the Kondo effect. The low temperature deviation from a Curie-Weiss law in the susceptibility and a step in the susceptibility, along with the excess specific heat, provide further evidence for the presence of magnetic correlations at low temperature in our single crystal samples of VSe$_2$. Deviations from a Curie-Weiss law may also be due to the presence of a Kondo screening of the magnetic moments on the magnetic impurities. It is to be noted that in another TMDC, 1T-TaS$_2$, an increase in resistivity with decreasing temperature was reported below 60~K\cite{Disalvo1977}. However, the magnetoresistance was always negative for the whole range of field, except for a small temperature range between 0.1 and 2.2~K where it was positive at low fields. This was interpreted to be the result of variable range hopping in the Anderson localized states\cite{Kobayashi1979}. In TMDCs generally a positive magnetoresistance is seen\cite{Ali2014,Zhou2016}. In our samples at low temperatures, the negative magnetoresistance from the Kondo effect and a linear positive magnetoresistance compete with each other, with the positive magnetoresistance finally dominating at higher fields. The fact that the Kondo effect disappears as the temperature rises, explains why only a positive magnetoresistance is seen at higher temperatures. The signatures of the Kondo effect seen in the low-temperature resistivity and magnetoresistance of our single crystal samples of VSe$_2$ are not widely reported in other TMDCs and this makes an interesting addition to the already diverse list of properties of the TMDCs.

\section*{Methods}

\subsection*{Polycrystalline powder and single crystal growth}

VSe$_2$ samples were first synthesized in polycrystalline form by reacting powders of the starting materials, vanadium (Alfa Aesar, 99.5\%) and selenium (Alfa Aesar, 99.999\%). Stoichiometric amounts of the powders were ground together and transferred into quartz tubes, which were evacuated and sealed. The polycrystalline mixture was then heat treated for several days at 750~$^{\circ}$C, followed by a second heating at temperatures in the range 815-825~$^{\circ}$C for several days. Single crystals of VSe$_{2}$ were grown by the chemical vapour transport method by using the polycrystalline VSe$_2$ powder as the starting material.  These powders were sealed in a quartz tube under vacuum and then kept in a temperature gradient using the procedure described in Refs.~\citenum{Bayard1976,Wehmeier1970,Vanbruggen1976}. Iodine was used as the transporting agent.

\subsection*{Sample characterization}

Phase composition analysis was carried out using a Bruker D5005 X-ray diffractometer using Cu K$\alpha_{1}$ and K$\alpha_{2}$ radiation ($\lambda_{\mathrm{K}\alpha 1} = 1.5406$~\AA~and $\lambda_{\mathrm{K}\alpha 2}  = 1.5444$~\AA). The powder X-ray diffraction patterns were collected at room temperature over an angular range of 8 to $110^{\circ}$ in 2$\theta$ with a step size in the scattering angle 2$\theta$ of 0.02$^{\circ}$, and a total scanning time of 17~hours. The analysis of the powder X-ray patterns was performed using the FullProf software suite\cite{Rodriguez1993}. Single crystal X-ray diffraction measurements were performed by mounting a small piece of single crystal measuring $12~\mu$m $\times 43~\mu$m $\times 126~\mu$m, cleaved from the bulk, on a Gemini R CCD X-ray diffractometer (Agilent Technologies). The data were collected using Mo K$\alpha$ radiation ($\lambda = 0.71073$~\r{A}) at room temperature, and were then indexed and integrated using CrysAlisPro (Agilent Technologies). The structure was refined using Shelx\cite{Sheldrick2008}. 398 reflections were measured, of which 81 were independent reflections. An extinction parameter was refined and the obtained value was 0.04(6). Attempts to fit the single crystal X-ray data using different models for the occupancies of the V and Se sites, showed that the best model for the refinements is one in which the occupancies of both V and Se sites are fixed. The refinement was made using isotropic atomic displacement for both V and Se sites. A Laue X-ray imaging system with a Photonic-Science Laue camera was used to investigate the quality of the crystals prepared. The composition of the single crystals was also determined using an energy dispersive X-ray spectrometer in a Zeiss SUPRA 55--VP scanning electron microscope.

\subsection*{Transport and Magnetic Measurements}

Resistivity and magnetoresistance measurements down to 2~K were performed using an AC transport measurement option in a Quantum Design Physical Property Measurement System (PPMS). DC resistivity measurements down to 0.5~K were performed using a DC transport measurement option in a PPMS with a Helium-3 insert. Electrical contacts were made to the sample in a four probe configuration with silver wires and conductive silver paint for the resistivity and magnetoresistance measurements. The Hall resistivity was measured using a five wire Hall measurement configuration in the AC transport option. Currents of 1 to 10~mA were used in the AC resistivity and magnetoresistance measurements, 0.1 to 0.5~mA for the low-temperature DC resistivity measurements, while in the Hall measurements currents of 0.05~mA were used. The angle dependence of the magnetoresistance was measured using the horizontal rotator option in the PPMS. The DC magnetization measurements were performed using a Quantum Design Magnetic Property Measurement System MPMS-5S superconducting quantum interference device (SQUID) magnetometer for temperatures between 2 and 300~K and in an MPMS magnetometer with an \textit{i}Quantum $^{3}$He system in the temperature range 0.5 to 2~K. Low temperature heat capacity measurements were made using a two-tau relaxation method in a Quantum Design PPMS equipped with a $^3$He insert.

\bibliography{References}

\section*{Acknowledgements}

We acknowledge the EPSRC, UK for funding this work (Grant Nos. EP/L014963/1 and EP/M028771/1). We wish to thank D.T. Adroja (ISIS facility, STFC, RAL, U.K.) for valuable discussions and suggestions.

\section*{Author contributions statement}

G.B. conceived of the project. M.C.H synthesized the powder and the crystals and performed the X-ray diffraction analysis. S.B. performed all the characterization measurements with M.R.L. and analysed the data with M.R.L. S.B. drafted the paper with significant contributions from all the authors.

\section*{Additional information}

\textbf{Supplementary information} accompanies this paper at http://www.nature.com/srep \\
\textbf{Competing financial interests} The authors declare no competing financial interests.
%
%

\end{document}